\documentclass[aps,prd,preprintnumbers,superscriptaddress,showpacs,floatfix]{revtex4}
\usepackage{amssymb}
\usepackage{amsmath}
\usepackage{graphicx}
\usepackage{bm}
\usepackage{subfigure}
\usepackage{natbib}

\begin{document}

\title{Transport Coefficients of Bulk Viscous Pressure in the 14-moment
approximation}
\date{\today }
\author{G.\ S.\ Denicol${}^{a}$, S.~Jeon${}^{a}$, and C.~Gale${}^{a,b}$}
\affiliation{$^{a}$Department of Physics, McGill University, 3600 University Street, Montreal,
Quebec, H3A 2T8, Canada }
\affiliation{$^{b}$Frankfurt Institute for Advanced Studies, Ruth-Moufang-Str.\ 1, D-60438
Frankfurt am Main, Germany}

\begin{abstract}
We compute the transport coefficients that appear in the fluid-dynamical
equations for the bulk viscous pressure and shear-stress tensor using the
14-moment approximation in the limit of small, but finite, masses.
In this limit, we are able to express all these coefficients in terms of
known thermodynamic quantities, such as the thermodynamic pressure, energy
density, and the velocity of sound. We explicitly demonstrate that the ratio
of bulk viscosity to bulk relaxation time behaves very differently, as a
function of temperature, than the ratio of shear viscosity to shear
relaxation time. We further explicitly compute, for the first time, the
transport coefficients that couple the bulk viscous pressure to the
shear-stress tensor and vice versa. The coefficient that couples bulk
viscous pressure to shear-stress tensor is found to be orders of magnitude
larger than the bulk viscosity itself, suggesting that bulk viscous pressure 
production owes more to this coupling than to the expansion rate of the system. 
\end{abstract}

\maketitle


\section{Introduction}

The effect of dissipation on the dynamics of the quark-gluon plasma (QGP)
created in ultrarelativistic heavy ion collisions has been widely
investigated by several authors \cite{review}. In the last 5 years, most of
the effort on this topic was concentrated in extracting the magnitude of the
shear viscosity of the QGP from experiment, i.e., in investigating how small
the shear viscosity to entropy density ratio actually is \cite{Romatschke1,
Luzum:2008cw, Song:2008hj, Song:2010aq, Song:2011hk, Schenke:2010rr,
Song:2010mg,Schenke:2011tv,
Song:2011qa,Schenke:2011qd,New,Niemi:2011ix,Niemi:2012ry}. In
parallel, while attempts were made to understand the behavior of the bulk
viscosity of hot and dense nuclear matter in general \cite%
{Ozvenchuk:2012kh,Huang:2011ez,Huang:2010sa,Dobado:2011qu,
Lu:2011df,Bluhm:2010qf,Chakraborty:2010fr,FernandezFraile:2008vu,NoronhaHostler:2008ju,Jeon:1995zm,Jeon:1994if,Florkowski:2014sfa}%
, and while some simulations of relativistic heavy ion collisions did
consider the effects of bulk viscous pressure on the dynamics of the QGP 
\cite%
{Paech:2006st,Torrieri:2008ip,Rajagopal:2009yw,Song:2009rh,Monnai:2009ad,Bozek:2009dw,Denicol:2010tr,Dusling:2011fd,Noronha-Hostler:2013gga,Denicol:2009am}%
, it is fair to write that the effects of bulk viscosity still remain to be
studied systematically in numerical simulations of heavy-ion collisions.

At very high temperatures, it is established that bulk viscosity of QCD
matter is much smaller than the shear viscosity, which led to the belief
that bulk viscosity would play a much smaller role when compared to shear
viscosity in the description of the bulk nuclear matter created in heavy ion collisions. However, one should still note
that, in the temperature region produced experimentally in heavy ion
collisions, the actual order of magnitude and temperature dependence of bulk
viscosity is unknown \cite{Moore:2008ws} and, in principle, it
can be large enough to affect the time evolution of bulk QCD matter.

When it comes to the behavior of the remaining transport coefficients that
appear in the equation of motion for the bulk viscous pressure, the
uncertainties are even larger. Basic features such as the behavior of the
bulk relaxation time, how the bulk viscous pressure couples to the
shear-stress tensor, among others, are not well understood even on a
qualitative level. Together with uncertainties on how to implement the
freezeout procedure when bulk viscous pressure is present \cite%
{Denicol:2009am,Denicol:2012yr,Pratt:2010jt}, this makes it more complicated
to establish concrete results about the influences of bulk viscosity on
heavy ion collision observables on a phenomenological level.

Therefore, it is relevant to study the aforementioned uncertainties first in
simpler systems, where they can be analyzed in detail. For instance, these
questions can be addressed in the framework of the relativistic Boltzmann
equation, where all transport coefficients that appear in the equations of
relativistic fluid dynamics are in principle known and can be explicitly
computed. Similar investigations were already done for shear-stress tensor
and heat flow \cite{heatflow}.

The equations of motion and transport coefficients of a relativistic fluid
can be derived from the relativistic Boltzmann equation using the 14-moment
approximation \cite{IS,DKR,DNMR}. The main equations of motion are the
conservation laws of energy, momentum, and charge, which follow directly
from the conservation of these quantities on a microscopic level. For the
sake of simplicity, in this paper we will always assume that the net charge
is always zero and that no charge diffusion takes place. Then, we only have
to solve the continuity equations for the energy-momentum tensor, $T^{\mu
\nu }$, 
\begin{equation*}
\partial _{\mu }T^{\mu \nu }=0\text{ }.
\end{equation*}%
The above equation is general and independent of the formalism employed to
derive fluid dynamics. The remaining 6 equations of motion that close this
system (9 equations, if one takes into account charge diffusion), i.e., the
time evolution equations for the shear-stress tensor and bulk viscous
pressure, are less general and must be derived within a certain framework.
In the case of the Boltzmann equation, a possible framework is the already
mentioned 14-moment approximation developed by Israel and Stewart (in
principle, it should be called 10-moment approximation, since we don't
include net charge density and diffusion 4-current, but we shall keep naming
it 14-moment nevertheless). In the case where only bulk viscous pressure and
shear-stress tensor are present, the 14-moment approximation leads to the
following relaxation-type equations \cite{DKR,DNMR} 
\begin{eqnarray}
\tau _{\Pi }\dot{\Pi}+\Pi  &=&-\zeta \theta -\delta _{\Pi \Pi }\Pi \theta
+\varphi _{1}\Pi ^{2}+\lambda _{\Pi \pi }\pi ^{\mu \nu }\sigma _{\mu \nu
}+\varphi _{3}\pi ^{\mu \nu }\pi _{\mu \nu }\;,  \label{intro_1} \\
\tau _{\pi }\dot{\pi}^{\left\langle \mu \nu \right\rangle }+\pi ^{\mu \nu }
&=&2\eta \sigma ^{\mu \nu }+2\pi _{\alpha }^{\left\langle \mu \right.
}\omega ^{\left. \nu \right\rangle \alpha }-\delta _{\pi \pi }\pi ^{\mu \nu
}\theta +\varphi _{7}\pi _{\alpha }^{\left\langle \mu \right. }\pi ^{\left.
\nu \right\rangle \alpha }-\tau _{\pi \pi }\pi _{\alpha }^{\left\langle \mu
\right. }\sigma ^{\left. \nu \right\rangle \alpha }  \notag \\
&&\;+\lambda _{\pi \Pi }\Pi \sigma ^{\mu \nu }+\varphi _{6}\Pi \pi ^{\mu \nu
}.  \label{intro_2}
\end{eqnarray}%
Here, $\Pi $ is the bulk viscous pressure and $\pi ^{\mu \nu }$ is the
shear-stress tensor. We further introduced the vorticity tensor $\omega
_{\lambda \rho }\equiv \left( \nabla _{\lambda }u_{\rho }-\nabla _{\rho
}u_{\lambda }\right) /2$, the shear tensor $\sigma _{\lambda \rho }\equiv
\nabla _{\left\langle \lambda \right. }u_{\left. \rho \right\rangle }$ and
the expansion scalar $\theta \equiv \nabla _{\mu }u^{\mu }$, with $u^{\mu }$
being the fluid 4-velocity, $\nabla _{\mu }=\Delta _{\mu }^{\nu }\partial
_{\nu }$ the projected spatial gradient. We used the notation $A^{\langle
\mu \nu \rangle }\equiv \Delta _{\alpha \beta }^{\mu \nu }A^{\alpha \beta }$%
, with $\Delta _{\alpha \beta }^{\mu \nu }\equiv (\Delta _{\alpha }^{\mu
}\Delta _{\beta }^{\nu }+\Delta _{\beta }^{\mu }\Delta _{\alpha }^{\nu
}-2/3\Delta ^{\mu \nu }\Delta _{\alpha \beta })/2$, $\Delta ^{\mu \nu
}=g^{\mu \nu }-u^{\mu }u^{\nu }$, and $g^{\mu \nu }=\mathrm{diag}(1,-1,-1,-1)
$. The quantities multiplying each one of the terms are their corresponding
transport coefficients, which are general functions of temperature (and
chemical potential, if any) that should be extracted by matching fluid
dynamics to the underlying microscopic theory, which here corresponds to the
relativistic Boltzmann equation.

The above equations complement the conservation laws of energy-momentum.
General formulas for the transport coefficients $\beta _{\Pi }$, $\delta
_{\Pi \Pi }$, $\lambda _{\Pi \pi }$, $\beta _{\pi }$, $\delta _{\pi \pi }$, $%
\tau _{\pi \pi }$, and $\lambda _{\pi \Pi }$ were obtained in Refs.~\cite%
{DKR,DNMR}, while expressions for $\varphi _{3}$, $\varphi _{6}$, and $%
\varphi _{7}$ were recently computed in Ref.~\cite{Molnar:2013lta}. However,
such transport coefficients have been expressed formally, with most of them
still not written in a convenient form to be implemented in fluid-dynamical
simulations of heavy ion collisions or to provide a qualitative
understanding of how these coefficients behave.

The coefficients $\delta _{\pi \pi }$, $\tau _{\pi \pi }$, and $\varphi _{7}$
do not vanish in the massless limit and, in this limit, can be trivially
related to the shear viscosity relaxation time, $\tau _{\pi }$, and the
thermodynamic pressure, $P_{0}$, \cite{DNMR,Molnar:2013lta},%
\begin{equation}
\delta _{\pi \pi }=\frac{4}{3}\tau _{\pi }\text{ },\text{ \ }\tau _{\pi \pi
}=\frac{10}{7}\tau _{\pi }\text{ },\text{ \ }\varphi _{7}=\frac{9}{70P_{0}}%
\text{ }.
\end{equation}%
The formulas above are simple enough and can be used as estimates for 
$\delta _{\pi \pi }$, $\tau _{\pi \pi }$, and $\varphi _{7}$ in simulations
of heavy ion collisions. As a matter of fact, the above expression for the
transport coefficient $\delta _{\pi \pi }$ is already employed in all
simulations of heavy ion collisions that include the shear-stress tensor. In
Ref.\ \cite{Niemi:2012aj} the coefficients $\tau _{\pi \pi }$ and $\varphi
_{7}$, and their corresponding nonlinear terms, were already included in
fluid-dynamical simulations of heavy ion collisions. So far, similar
expressions do not exist for the coefficients of bulk viscous pressure,
which makes introducing bulk viscosity in heavy ion simulations in a
complete and consistent way a more complicated task.

In this paper, we investigate the behavior of all coefficients associated to
bulk viscous pressure and shear stress tensor when the mass is small
(relative to the temperature), but is still finite. In this limit, we are
able to express such coefficients in terms of the bulk and shear relaxation
times as well as other thermodynamic variables. In summary, we find the
following approximate expressions for the transport coefficients related to
bulk viscous pressure,%
\begin{eqnarray}
\frac{\zeta }{\tau _{\Pi }} &=&14.55\times \left( \frac{1}{3}%
-c_{s}^{2}\right) ^{2}\left( \varepsilon _{0}+P_{0}\right) +\mathcal{O}%
\left( z^{5}\right) ,  \notag \\
\frac{\delta _{\Pi \Pi }}{\tau _{\Pi }} &=&\frac{2}{3}+\mathcal{O}\left(
z^{2}\ln z\right) \text{ },  \notag \\
\frac{\lambda _{\Pi \pi }}{\tau _{\Pi }} &=&\frac{8}{5}\left( \frac{1}{3}%
-c_{s}^{2}\right) +\mathcal{O}\left( z^{4}\right) \text{ },
\end{eqnarray}%
where $z\equiv m/T$, with $m$ being the mass of the particle and $T$ the
temperature. The transport coefficients related to the shear-stress tensor
become%
\begin{eqnarray}
\frac{\eta }{\tau _{\pi }} &=&\frac{\varepsilon _{0}+P_{0}}{5}+\mathcal{O}%
\left( z^{2}\right) ,  \notag \\
\frac{\delta _{\pi \pi }}{\tau _{\pi }} &=&\frac{4}{3}+\mathcal{O}\left(
z^{2}\right) \text{ },  \notag \\
\frac{\tau _{\pi \pi }}{\tau _{\pi }} &=&\frac{10}{7}+\mathcal{O}\left(
z^{2}\right) \text{ },  \notag \\
\frac{\lambda _{\pi \Pi }}{\tau _{\pi }} &=&\frac{6}{5}+\mathcal{O}\left(
z^{2}\ln z\right) \text{ },
\end{eqnarray}%
As mentioned, the expressions above are valid when the mass is small, i.e., $%
m/T\ll 1$. Such formul{\ae} at least provide some intuition on how such
coefficients behave and also on how they are related to other transport
coefficients. In this sense, they can be useful to understand the parametric
dependence of the fluid-dynamical transport coefficients on $\varepsilon
_{0} $, $P_{0}$, $c_{s}$, $\tau _{\pi }$, and $\tau _{\Pi }$, making it less
complicated to implement them in the description of the strongly interacting system created in heavy ion collisions.

This paper is organized as follows. Sections \ref{Fluid_dynamics}, \ref%
{Exact_equations}, and \ref{14_moment_approx} summarize the basic steps
required to derive relativistic fluid dynamics from the Boltzmann equation
using the 14-moment approximation. Section \ref{RelaxTime_BGK} briefly
explains the relaxation time approximation. In section \ref%
{Expansion_small_masses} we derive the main results of this paper, obtaining
approximate expressions for the fluid-dynamical transport coefficients in
the small, but finite, mass limit. In section \ref{Conclusion} we summarize
our results and make our conclusions. Throughout this paper, we use natural
units $\hbar =c=k_{B}=1$.

\section{Fluid dynamics and kinetic theory}

\label{Fluid_dynamics}

For the sake of simplicity, in this work we only consider the case of a
single-component gas of classical particles. The starting point is the
relativistic Boltzmann equation 
\begin{equation}
k^{\mu }\partial _{\mu }f_{\mathbf{k}}=C\left[ f\right] ,  \label{BE}
\end{equation}%
where $C\left[ f\right] $ is the collision term, and $k^{\mu }=(k^{0}=\sqrt{%
\mathbf{k}^{2}+m^{2}},\mathbf{k)}$ with $m$ being the mass of the particle
considered. We further employ the notation $f_{\mathbf{k}}\equiv f(x^{\mu
},k^{\mu })$. For the purpose of this paper, it will not be necessary to
specify the collision term.

The energy-momentum tensor $T^{\mu \nu }$ is expressed as a moment of the
single particle distribution function 
\begin{equation}
T^{\mu \nu }=\left\langle k^{\mu }k^{\nu }\right\rangle \;,
\end{equation}%
where we employ the following notation%
\begin{equation}
\left\langle \ldots \right\rangle \equiv \int dK\left( \ldots \right) f_{%
\mathbf{k}}.
\end{equation}%
Above, $dK\equiv g\,d^{3}\mathbf{k}/\left[ (2\pi )^{3}k^{0}\right] $ is the
Lorentz-invariant measure, with $g$ being the appropriate degeneracy factor.

As usual, we define the fluid 4-velocity $u^{\mu }$ as an eigenvector of the
energy-momentum tensor, $T^{\mu \nu }u_{\mu }=\varepsilon u^{\nu }$, where
the eigenvalue $\varepsilon $ is identified with the fluid local energy
density \cite{landau}. Then, we can tensor-decompose the conserved current
in terms of the four-velocity%
\begin{equation}
T^{\mu \nu }=\varepsilon \,u^{\mu }u^{\nu }-\Delta ^{\mu \nu }\left(
P_{0}+\Pi \right) +\pi ^{\mu \nu }\;.
\end{equation}%
The energy density $\varepsilon $, the shear-stress tensor $\pi ^{\mu \nu }$%
, and the sum of thermodynamic pressure, $P_{0}$, and bulk viscous pressure, 
$\Pi $, are defined by 
\begin{equation}
\varepsilon \equiv \left\langle (u_{\mu }k^{\mu })^{2}\right\rangle \,,\text{
}\pi ^{\mu \nu }\equiv \left\langle k^{\left\langle \mu \right. }k^{\left.
\nu \right\rangle }\right\rangle \,,\;P_{0}+\Pi \equiv -\frac{1}{3}%
\left\langle \Delta ^{\mu \nu }k_{\mu }k_{\nu }\right\rangle \;.
\end{equation}%
We define the local equilibrium distribution function as $f_{0\mathbf{k}%
}=\exp \left( -\beta _{0}\,u_{\mu }k^{\mu }\right) $, where $\beta _{0}=1/T$
is the inverse temperature and we already assumed a vanishing chemical
potential ($\mu =0$). The temperature is defined from the following macthing
condition 
\begin{equation}
\varepsilon \equiv \varepsilon _{0}=\left\langle (u_{\mu }k^{\mu
})^{2}\right\rangle _{0},
\end{equation}%
where $\langle \ldots \rangle _{0}\equiv \int dK\left( \ldots \right) f_{0%
\mathbf{k}}$.

The thermodynamic pressure and bulk viscous pressure are then defined by 
\begin{equation}
P_{0}=-\frac{1}{3}\,\langle \Delta ^{\mu \nu }k_{\mu }k_{\nu }\rangle
_{0}\,,\Pi =-\frac{1}{3}\,\left\langle \Delta ^{\mu \nu }k_{\mu }k_{\nu
}\right\rangle _{\delta }\;,
\end{equation}%
with $\left\langle \ldots \right\rangle _{\delta }\equiv \left\langle \ldots
\right\rangle -\left\langle \ldots \right\rangle _{0}$.

\section{Equations of motion for $\Pi$ and $\pi^{\mu \nu}$}

\label{Exact_equations}

The equations of motion for $\Pi $ and $\pi ^{\mu \nu }$ can be computed
exactly for a single component gas using 
\begin{eqnarray}
\dot{\Pi} &=&-\frac{1}{3}m^{2}\int dK\,\delta \dot{f}_{\mathbf{k}}\;,
\label{Exact1} \\
\dot{\pi}^{\left\langle \mu \nu \right\rangle } &=&\int dK\,k^{\left\langle
\mu \right. }k^{\left. \nu \right\rangle }\,\delta \dot{f}_{\mathbf{k}}\;,
\label{Exact3}
\end{eqnarray}%
where we defined $\delta f_{\mathbf{k}}=f_{\mathbf{k}}-f_{0\mathbf{k}}$. The
comoving time derivative of $\delta f$, $\delta \dot{f}\equiv u^{\mu
}\partial _{\mu }\delta f$, can then be simplified using the relativistic
Boltzmann equation (\ref{BE}) in the form 
\begin{equation}
\delta \dot{f}_{\mathbf{k}}=-\dot{f}_{0\mathbf{k}}-\frac{1}{u_{\mu }k^{\mu }}%
\,k^{\mu }\nabla _{\mu }f_{\mathbf{k}}+\frac{1}{u_{\mu }k^{\mu }}C[f]\;.
\end{equation}%
After some algebra, one obtains the following equations,%
\begin{align}
\dot{\Pi}+C& =-\left[ \left( \frac{1}{3}-c_{s}^{2}\right) \left( \varepsilon
_{0}+P_{0}\right) -\frac{2}{9}\left( \varepsilon _{0}-3P_{0}\right) -\frac{%
m^{4}}{9}I_{-2,0}\right] \theta   \notag \\
& -\left( 1-c_{s}^{2}\right) \Pi \theta +\frac{m^{4}}{9}\rho _{-2}\theta
+\left( \frac{1}{3}-c_{s}^{2}\right) \pi ^{\mu \nu }\sigma _{\mu \nu }+\frac{%
m^{2}}{3}\rho _{-2}^{\mu \nu }\sigma _{\mu \nu },  \label{Exact_Eq_1} \\
\dot{\pi}^{\left\langle \mu \nu \right\rangle }+C^{\mu \nu }& =2\left[ \frac{%
4}{5}P_{0}+\frac{1}{15}\left( \varepsilon _{0}-3P_{0}\right) -\frac{m^{4}}{15%
}I_{-2,0}\right] \sigma ^{\mu \nu }+2\pi ^{\lambda \left\langle \mu \right.
}\omega _{\left. {}\right. \lambda }^{\left. \nu \right\rangle }  \notag \\
& -\left( \frac{4}{3}\pi ^{\mu \nu }+\frac{m^{2}}{3}\rho _{-2}^{\mu \nu
}\right) \theta +\left( \frac{6}{5}\Pi -\frac{2}{15}m^{4}\rho _{-2}\right)
\sigma ^{\mu \nu }  \notag \\
& -\left( \frac{10}{7}\pi ^{\lambda \left\langle \mu \right. }+\frac{4}{7}%
m^{2}\rho _{-2}^{\lambda \left\langle \mu \right. }\right) \sigma _{\lambda
}^{\left. \nu \right\rangle }.  \label{Exact_Eq_2}
\end{align}%
In the equations above, we have already neglected all terms related to
particle diffusion. Also, irreducible moments of $\delta f_{\mathbf{k}}$ of
rank higher than 2 were omitted, since such terms will not contribute to the
fluid-dynamical equations and exactly vanish in the 14-moment approximation.
The equations above are similar to those obtained in Refs.\ \cite{DKR,DNMR},
the only difference being that they were derived for a fixed chemical
potential, $\mu =\dot{\mu}=0$.

As in Ref.\ \cite{DNMR}, we use the following notation for the collision
terms, 
\begin{eqnarray}
C &=&\frac{m^{2}}{3}\int dK\text{ }(u_{\alpha }k^{\alpha })^{-1}C\left[ f%
\right] \;,  \notag \\
C^{\mu \nu } &=&-\int dK\text{ }(u_{\alpha }k^{\alpha })^{-1}k^{\langle \mu
}k^{\nu \rangle }C\left[ f\right] \;.
\end{eqnarray}%
and for the irreducible moments of $\delta f$%
\begin{eqnarray}
\rho _{n} &=&\left\langle (u_{\alpha }k^{\alpha })^{n}\right\rangle _{\delta
}\text{ },  \notag \\
\rho _{n}^{\mu } &=&\left\langle (u_{\alpha }k^{\alpha })^{n}k^{\left\langle
\mu \right\rangle }\right\rangle \text{ },  \notag \\
\rho _{n}^{\mu \nu } &=&\left\langle (u_{\alpha }k^{\alpha
})^{n}k^{\left\langle \mu \right. }k^{\left. \nu \right\rangle
}\right\rangle \text{ }.  \label{moments_def}
\end{eqnarray}

From equations (\ref{Exact_Eq_1}) and (\ref{Exact_Eq_2}), it is already
possible to identify the ratio of bulk viscosity to bulk relaxation time, $%
\zeta /\tau _{\Pi }$, and of shear viscosity to shear relaxation time, $\eta
/\tau _{\pi }$, as the following thermodynamic quantities%
\begin{eqnarray}
\frac{\zeta }{\tau _{\Pi }} &\equiv &\beta _{\Pi }=\left( \frac{1}{3}%
-c_{s}^{2}\right) \left( \varepsilon _{0}+P_{0}\right) -\frac{2}{9}\left(
\varepsilon _{0}-3P_{0}\right) -\frac{m^{4}}{9}I_{-2,0}\text{ },
\label{Relavant1} \\
\frac{\eta }{\tau _{\pi }} &\equiv &\beta _{\pi }=\frac{4}{5}P_{0}+\frac{1}{%
15}\left( \varepsilon _{0}-3P_{0}\right) -\frac{m^{4}}{15}I_{-2,0}\text{ }.
\label{Relevant2}
\end{eqnarray}%
Here, we introduced the velocity of sound squared, $c_{s}^{2}=dP_{0}/d%
\varepsilon _{0}$, which, at zero chemical potential, is given by%
\begin{equation}
c_{s}^{2}=\frac{\varepsilon _{0}+P_{0}}{\beta _{0}I_{30}}\text{ },
\label{cs2}
\end{equation}%
and made use of the thermodynamic functions $I_{nq}$, 
\begin{equation}
I_{nq}=\frac{1}{\left( 2q+1\right) !!}\int dK\text{ }\left( u_{\mu }k^{\mu
}\right) ^{n-2q}\left( -\Delta _{\mu \nu }k^{\mu }k^{\nu }\right) ^{q}f_{0%
\mathbf{k}}\ .  \label{Inq}
\end{equation}%
Note that, in this notation, $I_{20}=\varepsilon _{0}$ and $I_{21}=P_{0}$.
For a classical gas, the functions above satisfy the following rules%
\begin{eqnarray}
\beta _{0}I_{nq} &=&I_{n-1,q-1}+\left( n-2q\right) I_{n-1,q}\text{ },  \notag
\\
I_{n+2,q} &=&m^{2}I_{nq}+\left( 2q+3\right) I_{n+2,q+1}\text{ }.
\label{Inq_rel}
\end{eqnarray}%
These expressions are very convenient and can lead to rather useful
relations, some of which being,%
\begin{equation}
I_{10}=\beta _{0}I_{21}\text{ },\text{ }I_{31}=\frac{\varepsilon _{0}+P_{0}}{%
\beta _{0}}\text{ },\text{ }m^{2}I_{00}=\varepsilon _{0}-3P_{0}\text{ },%
\text{ }I_{42}=\frac{\varepsilon _{0}+P_{0}}{\beta _{0}^{2}}\text{ }.\text{ }
\label{Inq_example}
\end{equation}%
The last expression is widely employed in simulations of heavy ion
collisions when computing the nonequilibrium single-particle distribution
function associated to the shear-stress tensor.

When written in terms of the velocity of sound, the expression found for $%
\zeta/\tau_\Pi$ is exactly the same as the one found in Refs.~\cite%
{DKR,KOIDE}. It should be noted that, even though the formula is the same,
the behavior of the velocity of sound as a function of mass and temperature
is not.

\section{14-moment approximation}

\label{14_moment_approx}

Obviously, equations (\ref{Exact_Eq_1}) and (\ref{Exact_Eq_2}) are not
closed in terms of $\Pi $ and $\pi ^{\mu \nu }$ and, consequently, cannot be
the final form of the fluid-dynamical equations. On the other hand, it is
well known that a closed set of equations can be obtained from Eqs.\ (\ref%
{Exact_Eq_1}) and (\ref{Exact_Eq_2}) using the 14-moment approximation for
the single-particle distribution function, first introduced by Israel and
Stewart \cite{IS}. The 14-moment approximation dictates that 
\begin{equation}
\frac{f_{\mathbf{k}}-f_{0\mathbf{k}}}{f_{0\mathbf{k}}}=\left[
E_{0}+B_{0}m^{2}+D_{0}u_{\mu }k^{\mu }-4B_{0}\left( u_{\mu }k^{\mu }\right)
^{2}\right] \Pi +\lambda _{n}n_{\alpha }k^{\alpha }+B_{2}\pi _{\alpha \beta
}k^{\alpha }k^{\beta }\;.  \label{14m}
\end{equation}%
The fifth term in the above expansion, $\lambda _{n}n_{\alpha }k^{\alpha }$,
is related to particle diffusion and will be dropped, i.e., $\lambda _{n}=0$%
. The coefficients $E_{0}$, $D_{0}$, $B_{0}$, and $B_{2}$ are known
functions of $m$, $T$, and $u_{\mu }k^{\mu }$ (and also of $\mu $, when it
is nonzero), and can be expressed in terms of the thermodynamic functions, $%
I_{nq}$, as%
\begin{eqnarray}
B_{2} &=&\frac{1}{2I_{42}}\text{ },  \notag \\
\frac{D_{0}}{3B_{0}} &=&-4\frac{I_{31}I_{20}-I_{41}I_{10}}{%
I_{30}I_{10}-I_{20}I_{20}}\equiv -C_{2}\text{ },  \notag \\
\frac{E_{0}}{3B_{0}} &=&m^{2}+4\frac{I_{31}I_{30}-I_{41}I_{20}}{%
I_{30}I_{10}-I_{20}I_{20}}\equiv -C_{1}\text{ },  \notag \\
B_{0} &=&\frac{1}{-3C_{1}I_{21}-3C_{2}I_{31}-3I_{41}-5I_{42}}\text{ }.
\label{14_Coeffs}
\end{eqnarray}

In the 14-moment approximation (without particle diffusion), all moments of $%
\delta f_{\mathbf{k}}$ can be trivially written in terms of $\Pi $ and $\pi
^{\mu \nu }$, 
\begin{eqnarray}
\rho _{-n} &=&\gamma _{n}^{\left( 0\right) }\Pi \text{ },  \notag \\
\rho _{-n}^{\mu \nu } &=&\gamma _{n}^{\left( 2\right) }\pi ^{\mu \nu }\text{ 
}.  \label{moments}
\end{eqnarray}%
The coefficients $\gamma _{n}^{\left( 0\right) }$ and $\gamma _{n}^{\left(
2\right) }$ are complicated functions of $\beta _{0}$ and $m$ and read%
\begin{eqnarray}
\gamma _{n}^{\left( 0\right) } &=&\left( E_{0}+B_{0}m^{2}\right)
I_{-n,0}+D_{0}I_{1-n,0}-4B_{0}I_{2-n,0}\text{ },  \label{gamma_coeffs1} \\
\gamma _{n}^{\left( 2\right) } &=&\frac{I_{4-n,2}}{I_{42}}\text{ }.
\label{gamma_coeffs2}
\end{eqnarray}%
Note that only the coefficients $\gamma _{n}^{\left( 0,2\right) }$ with $n=2$
appear in the exact equations of motion for $\Pi $ and $\pi ^{\mu \nu }$.

Therefore, after making use of the 14-moment approximation, the equations of
motion become%
\begin{align}
\dot{\Pi}+C& =-\frac{\zeta }{\tau _{\Pi }}\theta +\frac{m^{2}}{3}\nabla
_{\mu }\rho _{-1}^{\mu }-\left[ 1-c_{s}^{2}-\frac{m^{4}}{9}\gamma
_{2}^{\left( 0\right) }\right] \Pi \theta   \notag \\
& +\left[ \frac{1}{3}-c_{s}^{2}+\frac{m^{2}}{3}\gamma _{2}^{\left( 2\right) }%
\right] \pi ^{\mu \nu }\sigma _{\mu \nu },  \label{Bulk_Eq} \\
\dot{\pi}^{\left\langle \mu \nu \right\rangle }+C^{\mu \nu }& =2\frac{\eta }{%
\tau _{\pi }}\sigma ^{\mu \nu }+2\pi ^{\lambda \left\langle \mu \right.
}\omega _{\left. {}\right. \lambda }^{\left. \nu \right\rangle }-\left[ 
\frac{10}{7}+\frac{4}{7}m^{2}\gamma _{2}^{\left( 2\right) }\right] \pi
^{\lambda \left\langle \mu \right. }\sigma _{\lambda }^{\left. \nu
\right\rangle }  \notag \\
& -\left[ \frac{4}{3}+\frac{m^{2}}{3}\gamma _{2}^{\left( 2\right) }\right]
\pi ^{\mu \nu }\theta +\left[ \frac{6}{5}-\frac{2}{15}m^{4}\gamma
_{2}^{\left( 0\right) }\right] \Pi \sigma ^{\mu \nu }\;.  \label{Shear_Eq}
\end{align}%
The bulk and shear relaxation times, $\tau _{\Pi }$ and $\tau _{\pi }$,
respectively, come from applying the 14-moment approximation to the
linearized collision term, leading to $C=\Pi /\tau _{\Pi }$ and $C^{\mu \nu
}=\pi ^{\mu \nu }/\tau _{\pi }$. The relaxation times, $\tau _{\Pi }$ and $%
\tau _{\pi }$, arise from the collision term and, consequently, depend on
the choice of cross section employed. All the other transport coefficients
will depend on the cross sections only through the relaxation time, i.e.,
their ratio with the relaxation time is independent of the interaction
strength. This will always be the case when the 14-moment approximation is
employed.

By comparing with the fluid-dynamical equations shown in the introduction,
Eqs.\ (\ref{intro_1}) and (\ref{intro_2}), we identify the most important
transport coefficients. The coefficients appearing in the equation of motion
for $\Pi $, besides $\zeta /\tau _{\Pi }$, are 
\begin{eqnarray}
\frac{\delta _{\Pi \Pi }}{\tau _{\Pi }} &=&1-c_{s}^{2}-\frac{m^{4}}{9}\gamma
_{2}^{(0)}\text{ }, \\
\frac{\lambda _{\Pi \pi }}{\tau _{\Pi }} &=&\frac{1}{3}-c_{s}^{2}+\frac{m^{2}%
}{3}\gamma _{2}^{(2)}\text{ },
\end{eqnarray}%
while those appearing in the equation of motion for $\pi ^{\mu \nu }$,
besides $\eta /\tau _{\pi }$, are found to be%
\begin{eqnarray}
\frac{\delta _{\pi \pi }}{\tau _{\pi }} &=&\frac{4}{3}+\frac{1}{3}%
m^{2}\gamma _{2}^{(2)}\text{ }, \\
\frac{\tau _{\pi \pi }}{\tau _{\pi }} &=&\frac{10}{7}+\frac{4}{7}m^{2}\gamma
_{2}^{(2)}\text{ }, \\
\frac{\lambda _{\pi \Pi }}{\tau _{\pi }} &=&\frac{6}{5}-\frac{2}{15}%
m^{4}\gamma _{2}^{(0)}\text{ }.
\end{eqnarray}%
The remaining coefficients $\varphi _{i}$ did not appear in the above
equations since they originate solely from the nonlinear component of the
collision term. In this paper, we will not consider these terms.

In the massless limit, only the coefficients $\beta _{\Pi }$ and $\lambda
_{\Pi \pi }$ vanish, i.e., $\beta _{\Pi }$ $=\lambda _{\Pi \pi }=0$. The
remaining coefficients are finite in this limit and become%
\begin{eqnarray}
\lim_{m/T\rightarrow 0}\frac{\eta }{\left( \varepsilon _{0}+P_{0}\right)
\tau _{\pi }} &=&\frac{1}{5},  \label{BLABLA} \\
\lim_{m/T\rightarrow 0}\frac{\delta _{\pi \pi }}{\tau _{\pi }} &=&\frac{4}{3}%
,\text{ }\lim_{m/T\rightarrow 0}\frac{\tau _{\pi \pi }}{\tau _{\pi }}=\frac{%
10}{7}, \\
\lim_{m/T\rightarrow 0}\frac{\lambda _{\pi \Pi }}{\tau _{\pi }} &=&\frac{6}{5%
},\text{ }\lim_{m/T\rightarrow 0}\frac{\delta _{\Pi \Pi }}{\tau _{\Pi }}=%
\frac{2}{3}.
\end{eqnarray}%
The coefficient $\delta _{\Pi \Pi }$ is the only one related to bulk viscous
pressure that does not vanish. This is not a problem since this is the
coefficient multiplying the term $\Pi \theta $ in Eq.\ (\ref{Bulk_Eq}) and $%
\Pi $ itself vanishes as $m^{2}$ when $m/T\rightarrow 0$. Therefore, in the
massless limit both sides of Eq.\ (\ref{Bulk_Eq}) will vanish, reducing it
to the trivial equality $0=0$.

Note that both the shear and bulk coefficients depend on the same
coefficients, $\gamma _{2}^{(0)}$ and $\gamma _{2}^{\left( 2\right) }$.
Therefore, the coefficients of bulk viscous pressure, $\zeta $, $\tau _{\Pi
} $, $\delta _{\Pi \Pi }$, and $\lambda _{\Pi \pi }$, can be exactly related
to those of the shear-stress tensor, $\eta $, $\tau _{\pi }$, $\delta _{\pi
\pi }$, $\tau _{\pi \pi }$, and $\lambda _{\pi \Pi }$, as follows%
\begin{eqnarray}
\frac{\zeta }{\tau _{\Pi }} &=&\frac{5}{3}\frac{\eta }{\tau _{\pi }}%
-c_{s}^{2}\left( \varepsilon _{0}+P_{0}\right) \text{ }, \\
\frac{\delta _{\Pi \Pi }}{\tau _{\Pi }} &=&\frac{5}{6}\frac{\lambda _{\pi
\Pi }}{\tau _{\pi }}-c_{s}^{2}, \\
\frac{\lambda _{\Pi \pi }}{\tau _{\Pi }} &=&\frac{1}{3}-c_{s}^{2}+\frac{14}{9%
}\frac{\tau _{\pi \pi }}{\tau _{\pi }}-\frac{5}{3}\frac{\delta _{\pi \pi }}{%
\tau _{\pi }}\text{ }.
\end{eqnarray}

\section{Relaxation time approximation}

\label{RelaxTime_BGK}

In this section, we compute the collision integrals using the
Bhatnagar-Gross-Krook (BGK) approximation \cite{BGK}, also known as the
relaxation time approximation. In the BGK formalism the collision term is
assumed to be proportional to the nonequilibrium single-particle momentum
distribution function,%
\begin{equation}
C\left[ f\right] =-u_{\mu }k^{\mu }\frac{\delta f_{\mathbf{k}}}{\tau _{R}}%
\text{ },
\end{equation}%
where $\tau _{R}$ is an energy dependent relaxation time scale. Usually,
this scale is parametrized to be of a certain power of $u_{\mu }k^{\mu }$,%
\begin{equation}
\tau _{R}\left( u_{\mu }k^{\mu }\right) =\tau _{\mathrm{mfp}}\left( \frac{%
u_{\mu }k^{\mu }}{T}\right) ^{\lambda },
\end{equation}%
with $\tau _{\mathrm{mfp}}$ being a time scale proportional to the mean-free
path of the system and $\lambda $ a parameter that specifies the power of
energy. For the sake of convenience, we fix $\lambda =0$.

With this choice of approximation, the collision integrals appearing in
Eqs.\ (\ref{Bulk_Eq}) and (\ref{Shear_Eq}) can be easily solved, leading to 
\begin{equation}
C=\frac{1}{\tau _{\mathrm{mfp}}}\Pi \;,\text{ \ }C^{\mu \nu }=\frac{1}{\tau
_{\mathrm{mfp}}}\pi ^{\mu \nu }\;.
\end{equation}%
Thus, we find the bulk and shear relaxation times to be equal, $\tau _{\Pi
}=\tau _{\pi }=$ $\tau _{\mathrm{mfp}}$. Since the BGK approximation
introduces only one time scale to describe the relaxation of the collision
operator, it is natural to expect all the fluid-dynamical relaxation times
to be proportional to this scale. However, one should note that both
relaxation times are only equal because we set $\lambda =0$. For other
values of $\lambda $, $\tau _{\Pi }\neq \tau _{\pi }$, even though they
would both remain proportional to $\tau _{\mathrm{mfp}}$.

\section{Expansion for small masses}

\label{Expansion_small_masses}

Using the properties of modified Bessel functions of the second kind, the
transport coefficients derived in the previous section can be expanded in a
series of $z=m/T$. The corresponding Bessel function can be written in an
integral form as%
\begin{equation}
K_{n}\left( z\right) =\frac{\Gamma \left( 1/2\right) }{\Gamma \left(
n+1/2\right) }\left( \frac{z}{2}\right) ^{n}\int_{0}^{\infty }d\theta \text{ 
}\left( \sinh \theta \right) ^{2n}\exp \left[ -z\cosh \left( \theta \right) %
\right] \text{ },
\end{equation}%
where $\Gamma $ is the gamma function \cite{SangyongsBook}.

The thermodynamic integrals $I_{nq}$ previously introduced can be cast in a
similar form,%
\begin{equation}
I_{nq}\left( T,z\right) =\frac{T^{n+2}}{\left( 2q+1\right) !!}\frac{z^{n+2}}{%
2\pi ^{2}}\int_{0}^{\infty }d\theta \text{ }\left( \cosh \theta \right)
^{n-2q}\left( \sinh \theta \right) ^{2q+2}\exp \left( -z\cosh \theta \right) 
\text{ }.
\end{equation}%
It is straightforward to express all thermodynamic integrals in terms of
Bessel functions, its derivatives, and/or its integrals. For example, the
thermodynamic pressure and energy density can be written in terms of $K_{1}$
and $K_{2}$ as 
\begin{eqnarray}
P_{0}\left( T,z\right)  &=&I_{21}\left( T,z\right) =\frac{1}{2}P_{0}\left(
T,0\right) z^{2}K_{2}\left( z\right) \text{ }, \\
\varepsilon _{0}\left( T,z\right)  &=&I_{20}\left( T,z\right) =3P_{0}\left(
T,z\right) +\frac{1}{2}P_{0}\left( T,0\right) z^{3}K_{1}\left( z\right) 
\text{ },
\end{eqnarray}%
while the particle number density as%
\begin{equation}
n_{0}\left( T,z\right) =I_{10}\left( T,z\right) =\frac{1}{2}n_{0}\left(
T,0\right) \left[ zK_{1}\left( z\right) -z^{2}\frac{d}{dz}K_{1}\left(
z\right) \right] \text{ }.
\end{equation}

The Bessel function $K_{n}\left( z\right) $ can be expanded in terms of $z$
as \cite{SangyongsBook}%
\begin{eqnarray}
K_{n}\left( z\right)  &=&\frac{1}{2}\sum_{k=0}^{n-1}\left( -1\right) ^{k}%
\frac{\left( n-k-1\right) !}{k!}\left( \frac{z}{2}\right) ^{2k-n}  \notag \\
&&+\left( -1\right) ^{n+1}\sum_{k=0}^{\infty }\frac{1}{k!\left( n+k\right) !}%
\left( \frac{z}{2}\right) ^{n+2k}\left[ \ln \frac{z}{2}-\frac{1}{2}\psi
\left( k+1\right) -\frac{1}{2}\psi \left( n+k+1\right) \right] \text{ },
\end{eqnarray}%
with $\psi $ being the digamma function. Using this series expansion of $%
K_{n}\left( z\right) $, one can obtain the dominant terms when $z=m/T\ll 1$
for several thermodynamic quantities. For the transport coefficients of bulk
viscous pressure, the most relevant quantities are%
\begin{eqnarray}
\frac{\varepsilon _{0}\left( T,z\right) -3P_{0}\left( T,z\right) }{%
P_{0}\left( T,0\right) } &=&\frac{1}{2}z^{2}+\frac{1}{8}z^{4}\left( 2\ln
z+2\gamma -1-\ln 4\right) +\mathcal{O}\left( z^{6}\ln z\right) \text{ }, 
\notag \\
\frac{1}{3}-c_{s}^{2} &=&\frac{1}{36}z^{2}-\frac{5}{864}z^{4}+\mathcal{O}%
\left( z^{6}\ln z\right) \text{ },  \notag \\
2\pi ^{2}I_{-20} &=&\ln z+a+\mathcal{O}\left( z\right) \text{ },  \notag \\
\frac{2\pi ^{2}}{T}I_{-10} &=&1-b\text{ }z+\mathcal{O}\left( z^{2}\ln
z\right) \text{ },  \notag \\
m^{2}\gamma _{2}^{(2)} &=&\frac{1}{20}z^{2}+\mathcal{O}\left( z^{4}\right) 
\text{ }.  \label{Important}
\end{eqnarray}%
where $a$ and $b$ are integration constants that can be identified to be $%
a=0.86$ and $b=1.6$. Furthermore, the expansion coefficients appearing in $%
\delta f$ can be shown to be all of order $\mathcal{O}\left( z^{-2}\right) $,%
\begin{equation}
B_{0}(z)=\frac{3}{4}z^{-2}+\mathcal{O}\left( 1\right) \text{ },\text{ \ }%
D_{0}(z)=24z^{-2}+\mathcal{O}\left( 1\right) \text{ },\text{ \ }%
E_{0}(z)=-36z^{-2}+\mathcal{O}\left( 1\right) \text{ \ }.
\label{df_expanded}
\end{equation}

Substituting Eqs.\ (\ref{Important}) and (\ref{df_expanded}) into Eq.\ (\ref%
{gamma_coeffs1}) one then obtains%
\begin{equation}
m^{4}\gamma _{2}^{\left( 0\right) }=18z^{2}\ln z+\mathcal{O}\left(
z^{2}\right) \text{ }.  \label{gamma_final}
\end{equation}%
Next, using Eqs.\ (\ref{Important}), (\ref{df_expanded}), and (\ref%
{gamma_final}) we can calculate the lowest order $z$ dependence of the
transport coefficients that vanish when $z=0$, i.e., the coefficients $\zeta
/\tau _{\Pi }$ and $\lambda _{\Pi \pi }/\tau _{\Pi }$. We find that 
\begin{equation}
\frac{\zeta }{\tau _{\Pi }}\sim \mathcal{O}\left( z^{4}\right) \text{ },%
\text{ \ }\frac{\lambda _{\Pi \pi }}{\tau _{\Pi }}\sim \mathcal{O}\left(
z^{2}\right) \text{ }.
\end{equation}%
As already mentioned, the remaining coefficients do not vanish and,
consequently, $\beta _{\pi }/\left( \varepsilon _{0}+P_{0}\right) \sim
\delta _{\pi \pi }/\tau _{\pi }\sim \tau _{\pi \pi }/\tau _{\pi }\sim \delta
_{\Pi \Pi }/\tau _{\Pi }\sim \mathcal{O}\left( 1\right) $. It is
particularly interesting that the ratio $\zeta /\tau _{\Pi }$ behaves as $%
\mathcal{O}\left( z^{4}\right) $, since $\left( 1/3-c_{s}^{2}\right) $ and $%
(\varepsilon -3P_{0})/P_{0}$ behave as $\mathcal{O}\left( z^{2}\right) $ and 
$z^{4}I_{-2,0}$ behaves as $\mathcal{O}\left( z^{4}\ln z\right) $. However,
as it turns out, all the terms that contribute to order $\mathcal{O}\left(
z^{2}\right) $, $\mathcal{O}\left( z^{2}\ln z\right) $, and $\mathcal{O}%
\left( z^{4}\ln z\right) $ simply cancel in the expression for $\zeta /\tau
_{\Pi }$, leaving a lowest order contribution of $\mathcal{O}\left(
z^{4}\right) $, as already stated. Such behavior is not obvious and could
not be inferred simply by looking at Eq.\ (\ref{Relavant1}).

As shown in Eq.\ (\ref{Important}), $1/3-c_{s}^{2}$ is of order $\mathcal{O}%
\left( z^{2}\right) $. Therefore, if we are only interested in the transport
coefficients up to order $\mathcal{O}\left( z^{4}\right) $, all terms
proportional to $z^{2}$ can be replaced by $1/3-c_{s}^{2}$. In this case, $%
\zeta /\tau _{\Pi }$ will be proportional $\left( 1/3-c_{s}^{2}\right) ^{2}$
while $\lambda _{\Pi \pi }/\tau _{\Pi }$ will just be proportional to $%
\left( 1/3-c_{s}^{2}\right) $. More explicitly, it is possible to prove that%
\begin{eqnarray}
\frac{\zeta }{\tau _{\Pi }} &=&14.55\times \left( \frac{1}{3}%
-c_{s}^{2}\right) ^{2}\left( \varepsilon _{0}+P_{0}\right) +\mathcal{O}%
\left( z^{5}\right) ,  \label{One} \\
\frac{\lambda _{\Pi \pi }}{\tau _{\Pi }} &=&\frac{8}{5}\left( \frac{1}{3}%
-c_{s}^{2}\right) +\mathcal{O}\left( z^{4}\right) \text{ }.  \label{two}
\end{eqnarray}%
Note that, at least when $z=m/T$ is small, $\lambda _{\Pi \pi }$, which
describes the coupling of bulk viscous pressure to shear-stress tensor, is
much larger than the bulk viscosity, $\zeta $, itself. This might indicate
that the coupling to shear-stress tensor ($\lambda _{\Pi \pi }\pi ^{\mu \nu
}\sigma _{\mu \nu }$) can give a contribution to the build up of the bulk
viscous pressure that is comparable to the one originating from the
Navier-Stokes term ($-\zeta \theta $). In this sense, such coupling term
should be included when investigated the effects of bulk viscous pressure.

Using the relaxation time approximation, as introduced in the previous
section, one can use Eqs.\ (\ref{BLABLA}) and (\ref{One}) to relate $\zeta $
and $\eta $,%
\begin{equation}
\zeta =72.75\text{ }\eta \left( \frac{1}{3}-c_{s}^{2}\right) ^{2}+\mathcal{O}%
\left( z^{5}\right) \text{ }.  \label{One_2}
\end{equation}%
Qualitatively, this expression is similar to the known relation, $\zeta
=15\eta \left( 1/3-c_{s}^{2}\right) ^{2}$, found by Weinberg more than 40 years
ago \cite{Weinberg}. However, quantitatively we obtain a proportionally
factor relating $\zeta $ and $\eta \left( 1/3-c_{s}^{2}\right) ^{2}$ that is
almost five times larger than the one displayed in Ref.\ \cite{Weinberg}.
This is not a problem since, motivated by applications to cosmology,
Weinberg derived his formula for a material medium, with very short
mean-free times, interacting with radiation. This is, of course, a very
different system than the one considered in this paper and there is no
reason to expect our calculation to coincide with Weinberg's.

Note that the temperature dependence of $\zeta /\tau _{\Pi }$ is quite
different from the temperature dependence of $\eta /\tau _{\pi }$. While the
latter is just proportional (in the ultrarelativistic limit) to $%
(\varepsilon _{0}+P_{0})$, the former displays a more complicated behavior,
being proportional to $(1/3-c_{s}^{2})^{2}(\varepsilon _{0}+P_{0})$. In this
sense, the approximation $\zeta /\tau _{\Pi }\sim (\varepsilon _{0}+P_{0})$
is not so good, even though it naively appears to be the simplest parametric
expression for $\zeta /\tau _{\Pi }$ that guarantees causality \cite%
{Causality}. On the other hand, for the case of shear viscosity $\eta /\tau
_{\pi }\sim (\varepsilon _{0}+P_{0})$ works very well, at least in the $%
m/T<1$ limit.

All the other transport coefficients behave as a function of $z=m/T$ in the
following way%
\begin{eqnarray}
\frac{\beta _{\pi }}{\varepsilon _{0}+P_{0}} &=&\frac{1}{5}-\frac{1}{60}%
z^{2}+\mathcal{O}\left( z^{4}\ln z\right) ,  \notag \\
\frac{\delta _{\pi \pi }}{\tau _{\pi }} &=&\frac{4}{3}+\frac{1}{60}z^{2}+%
\mathcal{O}\left( z^{4}\ln z\right) \text{ },  \notag \\
\frac{\tau _{\pi \pi }}{\tau _{\pi }} &=&\frac{10}{7}+\frac{1}{35}z^{2}+%
\mathcal{O}\left( z^{4}\ln z\right) \text{ },  \notag \\
\frac{\delta _{\Pi \Pi }}{\tau _{\Pi }} &=&\frac{2}{3}-\left( 2\ln
z+3.015\right) z^{2}+\mathcal{O}\left( z^{3}\right) \text{ },  \notag \\
\frac{\lambda _{\pi \Pi }}{\tau _{\pi }} &=&\frac{6}{5}-\left( \frac{12}{5}%
\ln z+3.65\right) z^{2}+\mathcal{O}\left( z^{3}\right) \text{ }.
\label{three}
\end{eqnarray}%
Most of them are constant up to order $\mathcal{O}\left( z^{2}\right) $. The
coefficients $\delta _{\Pi \Pi }/\tau _{\Pi }$ and $\lambda _{\pi \Pi }/\tau
_{\pi }$ are constant up to $\mathcal{O}\left( z^{2}\ln z\right) $,
displaying a stronger dependence on $m/T$. Overall, we find it to be a good
approximation to set all these terms as constants. Such constant values can
be thought of as lower bounds for these coefficients, since most of them
will increase with increasing $m/T$.

\begin{figure}[th]
\centering
\subfigure[]{\includegraphics[width=0.48\textwidth]{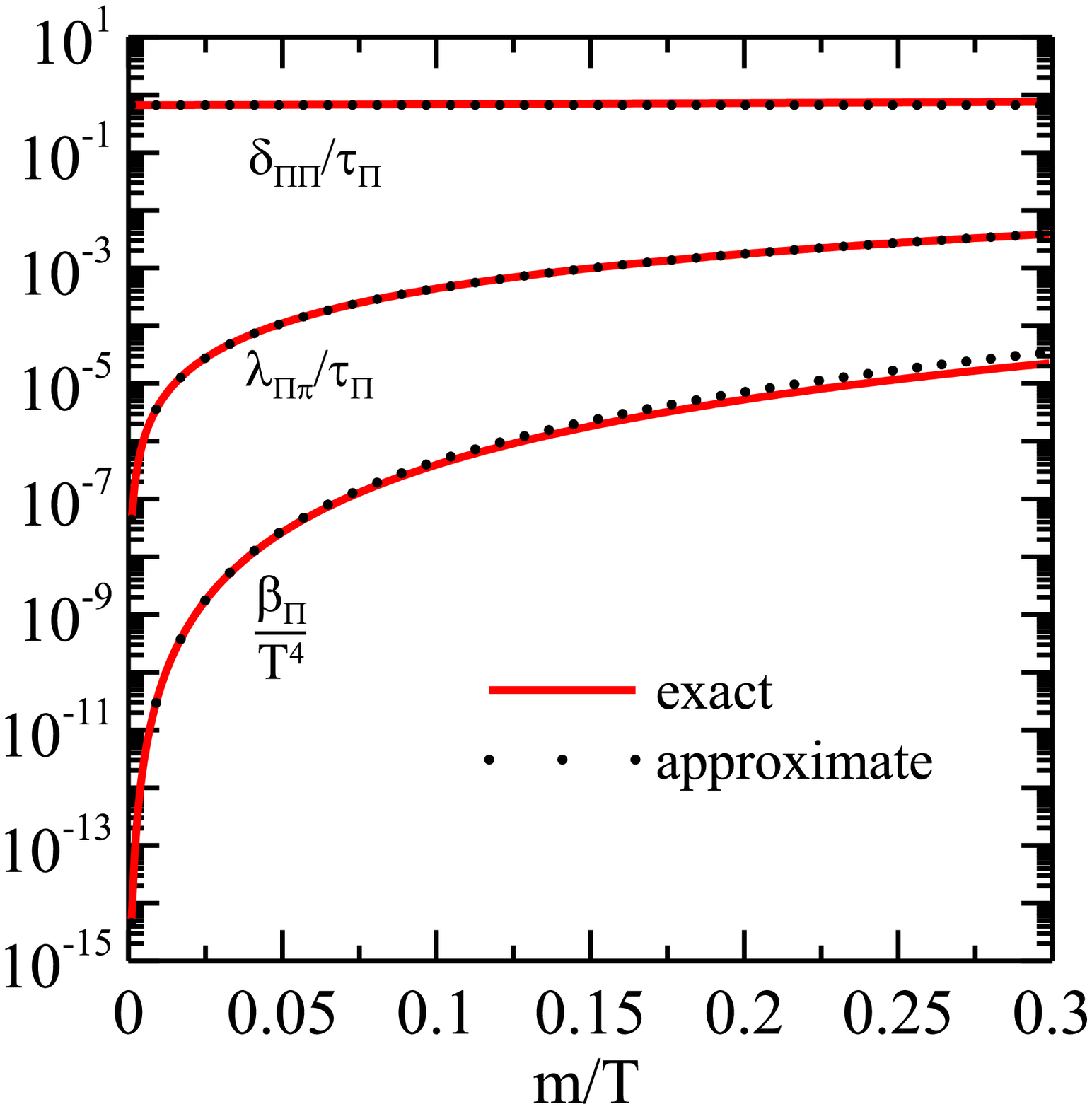}
\label{figEntropy:subfigure1}} \quad 
\subfigure[]{\includegraphics[width=0.48\textwidth]{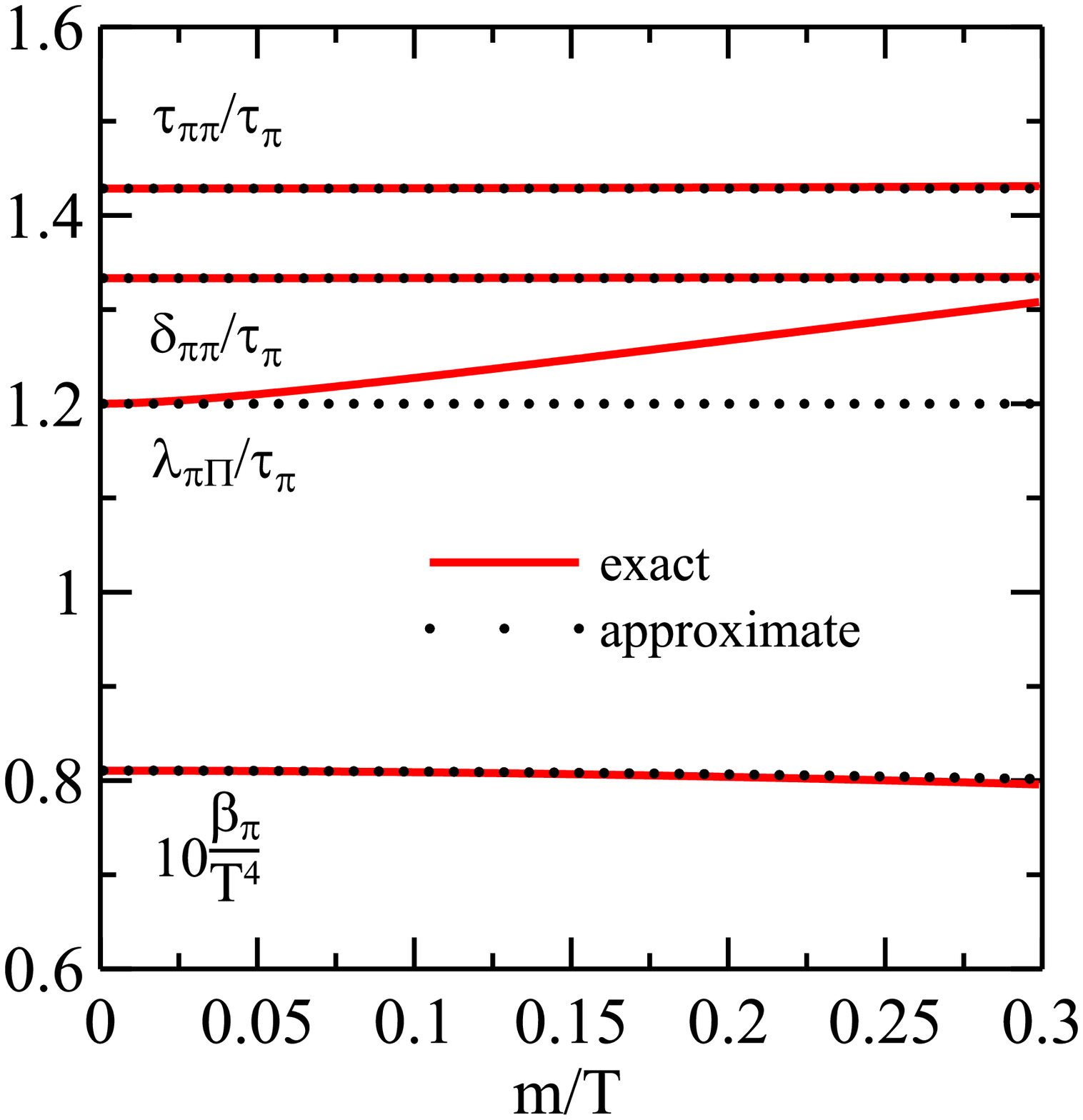}
\label{figEntropy:subfigure2}}
\caption{(Color online) In this figure we compare the approximate
expressions (circles) of the transport coefficients obtained in the main
text with their exact values (solid red lines), as a function of $mass/T$.
The left panel contains the transport coefficients appearing in the
equations of motion for the bulk viscous pressure while the right panel
contains the transport coefficients contained in the equations of motion for
the shear-stress tensor. }
\label{Comparison}
\end{figure}

The approximate expressions we found for the transport coefficients $\lambda
_{\pi \Pi }/\tau _{\pi }$ and $\lambda _{\Pi \pi }/\tau _{\Pi }$, which
describe the coupling between shear-stress tensor and bulk viscous pressure,
were the first to relate them to well-known thermodynamic quantities. In
this way, we can understand in a qualitative way, at least in the
ultrarelativistic limit, how such transport coefficients vary as we change
the velocity of sound. This simple behavior was unknown up to now.

The transport coefficient $\delta _{\Pi \Pi }/\tau _{\Pi }$ is somewhat
better understood. The term $\delta _{\Pi \Pi }\Pi \theta $, which appears
in the equation of motion for the bulk viscous pressure, is also commonly
written in a different, but equivalent, way as%
\begin{equation}
\frac{1}{2}\Pi \frac{\zeta T}{\tau _{\Pi }}\partial _{\mu }\left( \frac{\tau
_{\Pi }}{\zeta T}u^{\mu }\right) \text{ .}  \label{BLA}
\end{equation}%
Such term usually appears when deriving the equations of motion for $\Pi $
phenomenologically, via the second law of thermodynamics \cite{Muronga}. If
we use the expression $\zeta /\tau _{\Pi }$ found in this paper and
substitute it in Eq.\ (\ref{BLA}), it is possible to show that this term,
and the transport coefficient multiplying it, is also equivalent to the ones
obtained in this paper. However, if a different $\zeta /\tau _{\Pi }$ is
employed, one would then find that Eq.\ (\ref{BLA}) would lead to a
different and wrong expression for $\delta _{\Pi \Pi }/\tau _{\Pi }$.

In Fig.\ \ref{Comparison}, we compare the exact formula for these
coefficients with the approximate expressions just introduced in Eqs.\ (\ref%
{One}), (\ref{two}), and (\ref{three}). The approximate expressions for the
coefficients $\beta _{\Pi }$ and $\lambda _{\Pi \pi }/\tau _{\Pi }$ are
shown up to $\mathcal{O}\left( z^{5}\right) $ and $\mathcal{O}\left(
z^{4}\right) $, respectively, while only the constant parts (in $z$) of the
remaining coefficients are displayed. We see that the agreement between the
exact expressions and their approximate counterparts is reasonable up to $%
z\approx 0.3$, where largest difference observed could reach $20\%$ for $%
\delta _{\Pi \Pi }/\tau _{\Pi }$ and $\lambda _{\pi \Pi }/\tau _{\pi }$.

More precise, yet more complicated, expressions for these transport
coefficients can also be obtained. For the sake of completeness, we list
them below,%
\begin{eqnarray}
\frac{\beta _{\pi }}{\varepsilon _{0}+P_{0}} &=&\frac{1}{5}-\frac{3}{5}%
\left( \frac{1}{3}-c_{s}^{2}\right) +\mathcal{O}\left( z^{4}\ln z\right) 
\text{ },  \notag \\
\frac{\delta _{\pi \pi }}{\tau _{\pi }} &=&\frac{4}{3}+\frac{3}{5}\left( 
\frac{1}{3}-c_{s}^{2}\right) +\mathcal{O}\left( z^{4}\ln z\right) \text{ }, 
\notag \\
\frac{\tau _{\pi \pi }}{\tau _{\pi }} &=&\frac{10}{7}+\frac{36}{35}\left( 
\frac{1}{3}-c_{s}^{2}\right) +\mathcal{O}\left( z^{4}\ln z\right) \text{ }, 
\notag \\
\frac{\delta _{\Pi \Pi }}{\tau _{\Pi }} &=&\frac{2}{3}+\left[ 4-\frac{8}{9}%
\frac{\varepsilon _{0}-3P_{0}}{\left( \frac{1}{3}-c_{s}^{2}\right) \left(
\varepsilon _{0}+P_{0}\right) }\right] -108.89\times \left( \frac{1}{3}%
-c_{s}^{2}\right) +\mathcal{O}\left( z^{3}\right) \text{ },  \notag \\
\frac{\lambda _{\pi \Pi }}{\tau _{\pi }} &=&\frac{6}{5}+\frac{24}{5}\left[ 1-%
\frac{2}{9}\frac{\varepsilon _{0}-3P_{0}}{\left( \frac{1}{3}%
-c_{s}^{2}\right) \left( \varepsilon _{0}+P_{0}\right) }\right]
-127.02\times \left( \frac{1}{3}-c_{s}^{2}\right) +\mathcal{O}\left(
z^{3}\right) \text{ }.
\end{eqnarray}%
In the formulas above, the $z^{2}$ and $z^{2}\ln z$ dependences were removed
using the following relations%
\begin{eqnarray}
\frac{1}{3}-c_{s}^{2} &=&\frac{1}{36}z^{2}+\mathcal{O}\left( z^{4}\right) 
\text{ },  \notag \\
\frac{\left( \frac{1}{3}-c_{s}^{2}\right) \left( \varepsilon
_{0}+P_{0}\right) -\frac{2}{9}\left( \varepsilon _{0}-3P_{0}\right) }{\left( 
\frac{1}{3}-c_{s}^{2}\right) \left( \varepsilon _{0}+P_{0}\right) } &=&-%
\frac{1}{2}z^{2}\ln z-0.02536\times z^{2}+\mathcal{O}\left( z^{4}\ln
z\right) \text{ .}
\end{eqnarray}%
Most of these formulas are accurate up to order $\mathcal{O}\left( z^{4}\ln
z\right) $. Only the expressions for $\delta _{\Pi \Pi }/\tau _{\Pi }$ and $%
\lambda _{\pi \Pi }/\tau _{\pi }$, which always exhibit a stronger mass
dependence, are accurate only up to $\mathcal{O}\left( z^{3}\right) $.

It is important to remark that one can also find an approximate expression
for the momentum distribution function by expanding it in powers of $z$.
Using that%
\begin{eqnarray}
B_{0} &=&\frac{3}{4}\frac{1}{z^{2}}+\mathcal{O}\left( 1\right) \text{ }, 
\notag \\
D_{0} &=&24\frac{1}{z^{2}}+\mathcal{O}\left( 1\right) \text{ },  \notag \\
E_{0} &=&-36\frac{1}{z^{2}}+\mathcal{O}\left( 1\right) \text{ },
\end{eqnarray}%
and replacing these approximate expressions into Eq. (\ref{14m}), one can
show that%
\begin{equation*}
\frac{f_{\mathbf{k}}-f_{0\mathbf{k}}}{f_{0\mathbf{k}}}\approx \frac{4}{z^{2}}%
\left[ -36+24\frac{1}{T}u_{\mu }k^{\mu }-3\frac{1}{T^{2}}\left( u_{\mu
}k^{\mu }\right) ^{2}\right] \frac{\Pi }{\varepsilon _{0}+P_{0}}+\mathcal{O}%
\left( 1\right) \text{ }.
\end{equation*}%
This is an approximate and more convenient form for the single particle
distribution function, at least when the mass is relatively small compared
to the temperature.

\section{Conclusion}

\label{Conclusion}

In this paper we computed the transport coefficients that appear in the
fluid-dynamical equations for the bulk viscous pressure and shear-stress
tensor using the 14-moment approximation for a single component relativistic
dilute gas. The main results of this paper were computed in the limit of
small, but finite, mass where we showed it was possible to express all the
fluid-dynamical transport coefficients in terms of known thermodynamic
quantities, such as the thermodynamic pressure, energy density, and the
velocity of sound.

We explicitly demonstrated that the ratio of bulk viscosity to bulk
relaxation time behaves very differently, as a function of temperature, than
the ratio of shear viscosity to shear relaxation time. It was found that the
well known expression found by Weinberg, $\zeta =15\eta \left(
1/3-c_{s}^{2}\right) ^{2}$, for a system of radiation coupled to matter,
does not apply in general. In the limit of small $m/T$, one does in fact
find that $\zeta \sim \eta \left( 1/3-c_{s}^{2}\right) ^{2}$, but the
proportionality factor does not have to be $15$. Using the relaxation time
approximation, we actually found it to be $\sim 75$. When the mass is not
small, even the proportionality $\zeta \sim \eta \left( 1/3-c_{s}^{2}\right)
^{2}$ does not need to hold.

We further explicitly computed, for the first time, the transport
coefficients that couple the bulk viscous pressure to the shear-stress
tensor and vice versa. The coefficient that couples bulk viscous pressure to
shear-stress tensor is found to be orders of magnitude larger than the bulk
viscosity itself. This indicates that the bulk viscous pressure might be
produced in heavy ion collisions more due to its coupling to shear, than due
to the expansion rate of the system. So far, no simulations of heavy ion
collisions have ever taken into account the coupling that may exist between
different dissipative currents. For bulk viscous pressure, neglecting the
coupling to the shear-stress tensor might be a particularly bad
approximation.

All calculations in this paper were performed for a single component system
and neglect the effect of baryon chemical potential. This can be considered
a good feature, if one is just interested in the qualitative behavior of the
fluid-dynamical transport coefficients; especially on how they depend on
certain thermodynamic variables. In a future work, we plan to extend the
findings of this paper, considering a multi-component system and a
nonvanishing chemical potential. Then we shall have more quantitative results
and can check how well the qualitative expressions found in this paper can
describe more realistic systems.

\section{Acknowledgments}
G.~S.~Denicol acknowledges the
support of a Banting fellowship provided by the Natural Sciences and
Engineering Research Council of Canada. S.~Jeon acknowledges support by the Natural Sciences and
Engineering Research Council of Canada. C.~Gale acknowledges support from the Natural Sciences and Engineering Research Council of Canada, and by the 
Hessian initiative for excellence (LOEWE) through the Helmholtz International Center for FAIR (HIC for FAIR).

\end{document}